\documentclass[letter,twocolumn]{jpsj3}
\usepackage{bm}
\usepackage{braket}

\title{Reconstruction of chiral edge states in a magnetic Chern insulator}

\author{Ryo Ozawa\thanks{E-mail: ozawa@aion.t.u-tokyo.ac.jp}, Masafumi Udagawa, Yutaka Akagi, and Yukitoshi Motome}
\inst{Department of Applied Physics, University of Tokyo, Hongo 7-3-1, Bunkyo, Tokyo 113-8656, Japan
}

\abst{
We study reconstruction of the chiral edge states in a magnetic Chern insulator 
in the Kondo lattice model on a triangular lattice at 1/4 filling. 
In this state, the spin scalar chirality associated with a four-sublattice noncoplanar magnetic order assures the topological nature 
characterized by a nonzero Chern number. 
Performing a Langevin-based numerical simulation for finite-size clusters with the open boundary condition in one direction, 
we clarify how the magnetic and electronic properties are modulated near the edges of the system.
As a result, we find that the magnetic state near the edges is reconstructed to develop ferromagnetic spin correlations. 
At the same time, the electronic state is also modified and the chiral edge current is enhanced. 
We discuss this enhancement from the viewpoint of 
the electronic band structure for the gapless edge states.
}

\kword{
Chern insulator, chiral edge state, scalar chirality, 
Kondo lattice model, Langevin-based simulation}

\begin{document}
\maketitle

Since the discovery of the integer quantum Hall effect in two-dimensional electron systems under an external magnetic field, 
geometrical aspects of Bloch wave functions have attracted considerable attention
\cite{PhysRevLett.49.405, kohmoto1985topological}. 
The value of the Hall conductivity is determined by the Chern number, 
which characterizes the topological property of the Bloch wave functions in the two-dimensional magnetic Brillouin zone
\cite{kohmoto1985topological}.
Such a Chern insulating state accommodates peculiar edge states carrying an electric current in a specific direction, 
which are called the chiral edge states. 
These peculiar edge states are topologically protected, i.e., robust against perturbations. 

Recently, the Chern insulators have been found in several theoretical models even in the absence of an external magnetic field. 
For instance, the model on a honeycomb lattice proposed by Haldane exhibits the integer quantum Hall effect at zero field; 
a staggered magnetic flux penetrating the hexagonal plaquettes plays an important role in realizing the Chern insulating state
\cite{PhysRevLett.61.2015}. 
Other examples were found in the Kondo lattice model 
which describes the interplay between itinerant electrons and localized spins
\cite{PhysRevB.62.R6065,PhysRevLett.87.116801,PhysRevLett.101.156402,JPSJ.79.083711,
	PhysRevLett.105.226403,PhysRevLett.109.166405}. 
These Chern insulators are accompanied by peculiar noncoplanar spin textures with nonzero spin scalar chirality. 
The spin scalar chirality affects the motion of itinerant electrons through the spin Berry phase
which serves as an internal magnetic field
\cite{PhysRev.100.675}.
This leads to the integer quantum Hall effect with the quantization of the Hall conductivity and the chiral edge states. 

The magnetic Chern insulators in the spin-charge coupled systems have rich variety and controllability 
compared to the conventional integer quantum Hall systems in an applied magnetic field. 
Namely, as the magnetic Chern insulators are driven by an internal magnetic field from the spin texture, 
they have a variety depending on the electron filling, magnetic interactions, and lattice structures. 
Furthermore, since the spin degree of freedom couples to an external magnetic field through the Zeeman effect,
their properties can be controlled by the magnetic field.

On the other hand, in general, the spin-charge coupled systems change their properties near the surfaces of the system. 
The translational symmetry breaking gives rise to a significant reconstruction of the electronic and magnetic states near the surfaces. 
For instance, in perovskite manganites, the ferromagnetism stabilized by the spin-charge interplay is severely suppressed near the surfaces
\cite{PhysRevLett.81.1953}.
In the magnetic Chern insulators, however, the existence of chiral edge states brings new aspects to this problem.
On one hand, since the edge states are topologically protected, they are expected to survive the surface reconstruction.
On the other hand, the surface may affect the spin texture which sustains the nontrivial topological properties of the bulk system.
Accordingly, despite the topological protection, the fate of chiral edge states after a surface reconstruction is far from trivial.
It may also give an important viewpoint in the light of application to electronic devices, 
in which the surface may disturb the desired properties of chiral edge states.

In this Letter, 
we investigate the reconstruction of the chiral edge states in a magnetic Chern insulator realized in the Kondo lattice model.
We focus on a test ground of this problem: a scalar chiral ordered phase
appearing in the ground state in the Kondo lattice model on a triangular lattice at 1/4 filling
\cite{JPSJ.79.083711}.
We numerically optimize the magnetic and electronic structures to lower the total energy of the system with open edges 
by using an efficient and accurate algorithm based on the Langevin sampling.
By analyzing the optimized state in detail, we clarify spatial distributions of the 
spin correlation, spin scalar chirality, charge density, and electric current density near the edge. 
As a result, we find a drastic reconstruction of the magnetic state: 
strong ferromagnetic correlations are developed near the edges. 
This magnetic reconstruction affects the electronic nature of the chiral edge states: 
the total chiral edge current is enhanced and reaches almost twice compared to that without the reconstruction.
The enhancement of the chiral edge current is discussed in terms of the modulation of the band structure. 

We consider the Kondo lattice model on a triangular lattice, whose Hamiltonian is given by 
\begin{equation}
\hat{\mathcal H} = -t\sum_{\braket{l,m},s}(\hat{c}^\dagger_{ls}\hat{c}_{ms} 
+ \mathrm{h.c.}) - J_{\rm H}\sum_{l}\hat{\bm{s}}_l\cdot {\bm{S}_l}.
\label{eq:H}
\end{equation}
Here, the first term represents the kinetic energy of itinerant electrons, 
where $\hat{c}_{ls}$ ($\hat{c}_{ls}^\dagger$) is the annihilation (creation) operator of an itinerant electron 
with spin $s = \uparrow,\downarrow$ at $l$th site. 
The sum is taken over the nearest-neighbor sites on a triangular lattice. 
Hereafter, we take the transfer integral $t=1$ as an energy unit.
The second term denotes the on-site interaction between localized spins and itinerant electrons, 
where $J_{\rm H}$ represents the Hund's coupling constant. 
${\bm{S}_l}$ and ${\bm{\hat{s}}_l = \sum_{s,s'}\hat{c}^\dagger_{ls}\bm{\sigma}_{ss'}\hat{c}_{ls'}}$ 
represent the localized spin and itinerant electron spin operator at $l$th site, respectively
($\bm{\sigma}$ is the vector representation of the Pauli matrix).
In the following calculations, we assume $\bm{S}_l$ to be a classical vector with $|{\bm{S}}_l|=1$.

\begin{figure}[!h]
\begin{center}
\includegraphics[width=8cm]{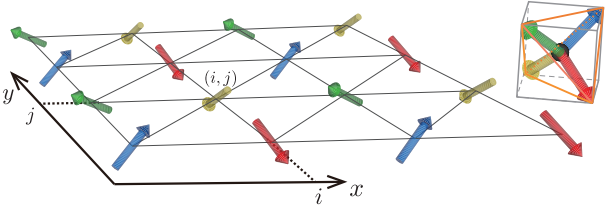}
\caption{
(Color online) 
Schematic picture of the magnetic structure in the four-sublattice scalar chiral ordered phase in the model in Eq.~(\ref{eq:H}). 
Four spins in the four-sublattice structure are oriented 
in the directions from the center to the four vertices of a tetrahedron (see the inset).
In the calculations, the open (periodic) boundary condition is imposed in the $x$ ($y$) direction. 
\label{fig:4sublattice}
}
\end{center}
\end{figure}

Magnetic Chern insulating phases were recently found as the stable ground state in the model in Eq.~(\ref{eq:H}) near 1/4 and 3/4 filling
\cite{PhysRevLett.101.156402,JPSJ.79.083711}. 
For both cases, the magnetic structure is given by the four-sublattice noncoplanar order shown in Fig.~\ref{fig:4sublattice}.
In this ordered state, the scalar chirality, $(\bm{S}_l \times \bm{S}_m) \cdot \bm{S}_n$, 
defined for three spins on each smallest triangle ($l$, $m$, and $n$ are in the counterclockwise direction), takes a uniform nonzero value. 
In the four-sublattice ordered phase with scalar chiral ordering,
the electronic spectrum is composed of four separated bands with the nonzero Chern number $-1, 1, 1, $ and $ -1$, from the top to bottom.
Hence, this system becomes a Chern insulator for $J_{\rm H} \gtrsim 0.7$ ($> 0$), if the Fermi level is in the energy gap of 1/4 (3/4) filling.

We numerically study the reconstruction of the chiral edge states in the Chern insulating phase at 1/4 filling. 
To obtain the equilibrium spin configuration, we adopt the recently-proposed Langevin-based method 
\cite{PhysRevB.88.235101}. 
In this method, the partition function $Z$ of the system is written by traces over itinerant electrons and classical localized spins as, 
\begin{equation}
Z = {\mathrm{Tr}}_{\{ \hat{c}_{ls}^\dagger, \hat{c}_{ls}\}}{\mathrm{Tr}}_{\{\bm{S}_l\}} 
e^{-\beta (\hat{\mathcal{H}}-\mu \hat{N})} 
= {\mathrm{Tr}}_{\{\bm{S}_l\}} e^{-\beta
F_{\rm el}(\{\bm{S}_l\})},
\label{eq:PF}
\end{equation}
where $\mu$ is the chemical potential, $\beta$ is the inverse temperature, 
and $\hat{N}$ is the total number operator for itinerant electrons. 
We set the Boltzmann constant $k_{\rm B}=1$. 
Here, $F_{\rm el}(\{\bm{S}_l\})$ is the free energy for a given configuration of localized spins $\{\bm{S}_l\}$, which is obtained as
\begin{equation}
F_{\rm el}(\{\bm{S}_l\}) = \int \rho(\varepsilon)
f_{\rm el}(\varepsilon) d\varepsilon.
\label{eq:FE}
\end{equation}
Here, $\rho(\varepsilon)$ is the density of states obtained from $\hat{\mathcal H}$ 
for the spin configuration $\{\bm{S}_l\}$, and $f_{\rm el}(\varepsilon) = -\beta^{-1} \log\{1 + \exp[-\beta(\varepsilon - \mu)]\}$.
In the Langevin-based method, the density of states $\rho(\varepsilon)$ in Eq.~(\ref{eq:FE}) is estimated by the kernel polynomial method
\cite{RevModPhys.78.275} 
and $\mathrm{Tr}_{\{\bm{S}_l\}}$ in Eq.~(\ref{eq:PF}) is replaced by the sum 
over spin configurations generated by the Langevin equation: 
\begin{equation}
\phi_l(\tau + \Delta \tau) 	= \phi_l(\tau) - \frac{\partial F_{\rm el} }{\partial \phi_l} \Delta \tau
					+ \sqrt{2\Delta \tau \beta^{-1}}\eta_l(\tau).
\label{eq:LE}
\end{equation}
Here, $\phi_l$ represents the polar and azimuth angles of the localized spin $\bm{S}_l$, $\tau$ represents a fictions time, 
and $\eta_l(\tau)$ in Eq.~(\ref{eq:LE}) is the uncorrelated Gaussian random variable with unit variance. 
The gradient of $F_{\rm el} $ is calculated by means of the automatic differentiation. 

We apply the Langevin-based method to the system with open (periodic) boundary condition in the $x$ ($y$) direction 
(see Fig.~\ref{fig:4sublattice}). 
In the following, we show the results for the system with $L_x\times L_y$ = 34 $\times$ 34 sites. 
We also calculated the systems from $L_x\times L_y = 24 \times 24$ to $48 \times 48$ and 
found that the finite-size effects are not relevant to the following results.  
We vary the value of the Hund's coupling $J_{\rm H}$ from 1.3 to 3.0,  
while changing the chemical potential $\mu$ to adjust 
the electron filling, $\sum_{l,s} \langle c_{ls}^\dagger c_{ls} \rangle/(2L_x L_y)$, to be 1/4.
In the kernel polynomial method, we expand the density of states $\rho(\varepsilon)$ by 1000 Chebyshev polynomials.
We calculate the Chebyshev moments of $\rho(\varepsilon)$ by using the complete basis set of $\hat{\cal H}$, 
rather than the stochastic estimation in the previous study
\cite{PhysRevB.88.235101}.
This allows to estimate Eq.~(\ref{eq:FE})  accurately in the large $\beta$ (low temperature) limit. 
In the Langevin equation in Eq.~(\ref{eq:LE}), we set the time interval $\Delta \tau = 0.5$.

To obtain the optimized state in the low temperature limit, 
we perform the Langevin-based simulation by omitting the last term in  Eq.~(\ref{eq:LE}) and setting $\beta=10^5$ in Eq.~(\ref{eq:FE}). 
We choose the four-sublattice ordered state as the initial state.
We also tried two other simulations:
one is to decrease the temperature gradually starting from a random spin configuration at high temperature (simulated annealing), 
and the other is to gradually lower the temperature from the vicinity of critical temperature of 
scalar chiral order ($\beta \simeq 10^2$~[\citen{PhysRevLett.105.266405}]) starting from the four-sublattice ordered state. 
However, with these two methods, the system turned out to be trapped at a metastable state with a higher energy, 
compared to that by the method we adopted.

In order to characterize the reconstruction of magnetic and electronic states near the edges,
we calculate the following quantities:
nearest-neighbor spin correlation $S_\parallel(i)$,
spin scalar chirality  $\chi(i)$,
local charge density $n(i)$, and
local electric current density $j_\parallel(i)$.
$S_\parallel (i)$ and $j_\parallel (i)$ are measured on the bonds parallel to the edges 
(along the $y$ direction), and all the quantities are obtained by averaging in the $y$ direction. 
The definitions are given as follows: 
\begin{align}
S_{\parallel}(i) &=\frac{1}{L_y}\sum_j \bm{S}_{(i,j)}\cdot\bm{S}_{(i,j+1)},
\label{eq:SS}\\
\chi(i) &= \frac{1}{2L_y}\sum_j\left(\kappa^\bigtriangleup_{(i,j)}+\kappa^\bigtriangledown_{(i,j)}\right),
\label{eq:chi}\\
n(i) &= \frac{1}{L_y}\sum_j\sum_{s=\uparrow,\downarrow}
\Braket{\hat{c}_{(i,j)s}^\dagger\hat{c}_{(i,j)s}},
\label{eq:n}\\
j_{\parallel}(i)&= \frac{1}{2{\rm i}L_y}\sum_j\sum_{s=\uparrow,\downarrow}
\Braket{\hat{c}_{(i,j),s}^\dagger\hat{c}_{(i,j+1),s} - \mathrm{h.c.}}.
\label{eq:j}
\end{align}
Here, $i$ and $j$ denote the $x$ and $y$ coordinate, 
taking $0\leq i\leq L_x-1$ and $0\leq j\leq L_y-1$, respectively (see Fig.~\ref{fig:4sublattice}).
The sum over $j$ are taken from $j$ = 0 to $L_y-1$. In Eq.~(\ref{eq:chi}), 
$\kappa^\bigtriangleup$ and $\kappa^\bigtriangledown$ denote the scalar chiralities for upward and downward triangles,  
defined  as
\begin{align*}
\kappa^\bigtriangleup_{(i,j)}     &= \left(\bm{S}_{(i,j)}\times\bm{S}_{(i+1,j   )}\right)\cdot\bm{S}_{(i+1, j+1)},\\ 
\kappa^\bigtriangledown_{(i,j)} &= \left(\bm{S}_{(i,j)}\times\bm{S}_{(i+1,j+1)}\right)\cdot\bm{S}_{(i, j+1)}, 
\end{align*}
respectively.
We note that in the four-sublattice magnetic ordered state in Fig.~\ref{fig:4sublattice}, the scalar chirality takes a uniform value: 
$\kappa^\bigtriangleup_{(i,j)}=\kappa^\bigtriangledown_{(i,j)}= 4/(3\sqrt{3})\sim 0.77$ for any $(i,j)$. 

\begin{figure}[!h]
\begin{center}
\includegraphics[width = 7cm]{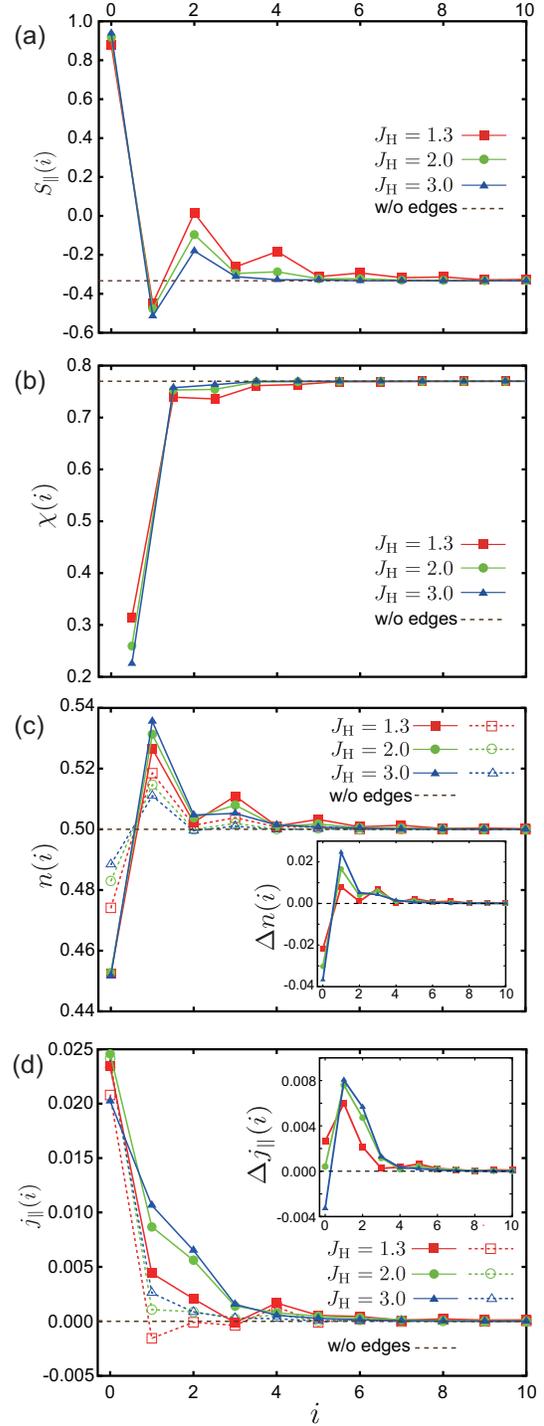} 
\caption{
(Color online)
(a) Neighboring spin correlation along the edge, $S_\parallel(i)$,
(b) spin scalar chirality $\chi(i)$,
(c) local charge density $n(i)$,
(d) electric current density along the edge, $j_\parallel(i)$,
as functions of the distance from the edge, $i$, 
calculated for the optimized state by the Langevin-based simulation for $L_x \times L_y = 34 \times 34$.
These quantities are obtained for 1/4 filling in the low-temperature limit;
the chemical potential is set at $\mu$ = -1.87, -2.30, and -3.20 for $J_{\rm H}$ = 1.3, 2.0, and 3.0, respectively. 
The dashed horizontal lines represent the bulk values. 
In (c) and (d), the data with dotted lines represent $n^{\rm 4-sub}(i)$ and $j^{\rm 4-sub}_\parallel(i)$. 
In the insets of (c) and (d), we plot $\Delta n(i) = n(i) - n^{\rm 4-sub}(i)$ and $\Delta j_\parallel(i) = j_{\parallel}(i) - j_\parallel^{\rm 4-sub}(i)$, respectively.
\label{fig:sp_modu}} 
\end{center}
\end{figure}

Figure~\ref{fig:sp_modu} shows the results obtained by the Langevin-based simulation. 
For comparison, the values in the bulk system are shown by the dashed horizontal lines. 
For the electronic properties in Figs.~\ref{fig:sp_modu}(c) and \ref{fig:sp_modu}(d), 
we also plot the data for the system with open edges in the perfectly-ordered four-sublattice chiral state; 
even when neglecting the reconstruction of the magnetic structure, $n(i)$ and $j_\parallel(i)$ deviate from the bulk values, 
because of the translational symmetry breaking by the edges. 
We denote these quantities as $n^{\rm 4-sub}(i)$ and $j_\parallel^{\rm 4-sub}(i)$. 
We will discuss these contributions separately from those by the reconstruction of the edge states.

As shown in Fig.~\ref{fig:sp_modu}(a), the spin correlation is drastically modulated near the edges. 
At the edge, $S_{\parallel}(0)$ takes $\sim 1$ for all the values of $J_{\rm H}$, 
indicating that ferromagnetic correlations are developed at the edge. 
The deviation from the bulk value oscillates as a function of $i$ and penetrates into the bulk for several layers.
The penetration depth becomes larger for smaller $J_{\rm H}$.
Corresponding to the ferromagnetic spin correlations developed at the edges, 
the spin scalar chirality is largely suppressed at the edge, as shown in Fig.~\ref{fig:sp_modu}(b).
However, in contrast to $S_\parallel(i)$, the change of $\chi(i)$ is rather limited near the edge;
for $i\geq1$, the change from the bulk value $4/(3\sqrt{3})$ is small. 
This is presumably because the scalar chirality is the high-order correlation in terms of spin. 

The change of magnetic structure affects the electronic states considerably. 
As shown in Fig.~\ref{fig:sp_modu}(c), 
the charge density $n(i)$ shows oscillating behavior near the edge.
This spatial modulation of $n(i)$ can be separated into two contributions:
one is due to the translational symmetry breaking by the edge and 
the other is from the coupling to the oscillating magnetic structure in Fig.~\ref{fig:sp_modu}(a). 
The former contribution corresponds to $n^{\rm 4-sub}(i)$, 
which is shown by  the data connected by the dotted lines in the main panel of Fig.~\ref{fig:sp_modu}(c).
The period of the oscillation is estimated as $\sim2$. 
This period is common to $S_\parallel(i)$ in Fig.~\ref{fig:sp_modu}(a), 
indicating that electronic oscillation from the translational symmetry breaking affects magnetic properties.
The latter contribution is extracted by substituting $n^{\rm 4-sub}(i)$ from $n(i)$;
we plot $\Delta n(i)= n(i) - n^{\rm 4-sub}(i)$ in the inset of Fig.~\ref{fig:sp_modu}(c).
As clearly shown in the result, the magnetic reconstruction leads to a substantial change in the charge density. 
The change becomes larger for larger $J_{\rm H}$.
At $J_{\rm H}=3.0$, more than half of the spatial modulation of $n(i)$ is ascribed to the latter contribution 
from the magnetic reconstruction. 

The changes of spin correlation and charge density follow the same tendency 
as those in the phase diagram in the bulk limit obtained in the previous study
\cite{JPSJ.79.083711}. 
The phase diagram shows that hole doping to the 1/4-filling chiral phase leads to the ferromagnetic phase.
Accordingly, the ferromagnetic layer is formed with the reduction of electron density, 
as shown in Figs.~\ref{fig:sp_modu}(a) and \ref{fig:sp_modu}(c).

As shown in Fig.~\ref{fig:sp_modu}(d), 
the current density $j_\parallel(i)$ also shows a characteristic spatial modulation near the edge. 
$j_\parallel(i)$ shows a maximal value at $i=0$, and slowly decays into the bulk. 
The decay is slower for smaller $J_{\rm H}$.
In contrast to $S_\parallel(i)$ and $n(i)$, $j_\parallel(i)$ does not show oscillating behavior.
The monotonic decay is rather analogous to that of $\chi(i)$, but the decay is much slower.
The spatial modulation of $j_{\parallel}(i)$ can also be separated into two contributions, as discussed for $n(i)$ above.
The contribution from the translational symmetry breaking, $j_\parallel^{\rm 4-sub}$, 
is shown by the data connected by the dotted lines in the main panel of Fig.~\ref{fig:sp_modu}(d), 
while the inset of Fig.~\ref{fig:sp_modu}(d) shows a contribution from the magnetic reconstruction, 
$\Delta j_\parallel(i) =j_\parallel(i) - j_\parallel^{\rm 4-sub}(i)$. 
As shown in Fig.~\ref{fig:sp_modu}(d),  
$j^{\rm 4-sub}_\parallel(i)$ is rather localized near the edge and rapidly decays into the bulk. 
On the other hand, $\Delta j_\parallel(i)$ shows nonmonotonic spatial dependence. 
While $\Delta j_\parallel(i=0)$ is rather small and even turns to negative for $J_{\rm H}\gtrsim 2.0$, 
$\Delta j_\parallel(i)$ has a broad peak at $i=1-2$. 
Furthermore, it is overall positive for $i \geq 1$, i.e., the reconstruction increases the net edge current. 

The nonmonotonic spatial dependence of $\Delta j_\parallel(i=0)$ is understood as follows. 
Suppose a perfect ferromagnetic configuration is stabilized for several layers from the edge due to the magnetic reconstruction. 
In this case, the system can be practically considered as a ``junction" of the ferromagnetic strip and the four-sublattice ordered bulk system. 
As a result, the chiral edge state appears at the boundary of the two subsystems, 
rather than the outer edge of the ferromagnetic strip, and accordingly, the edge current flows at the boundary. 
In our results, the outermost layer at $i=0$ becomes almost ferromagnetic, as shown in Fig.~\ref{fig:sp_modu}(a), 
approximately considered as a single-layer ferromagnetic strip. 
This picture might explain the enhancement of $\Delta j_{\parallel}(i)$ at inner layers, mostly at $i=1$ and $2$; namely, 
the enhancement is interpreted as the chiral edge current induced 
at the new ``boundary" between ferromagnetic skin and four-sublattice ordered bulk.

\begin{figure}[!h]
\begin{center}
\includegraphics[width = 7.2cm]{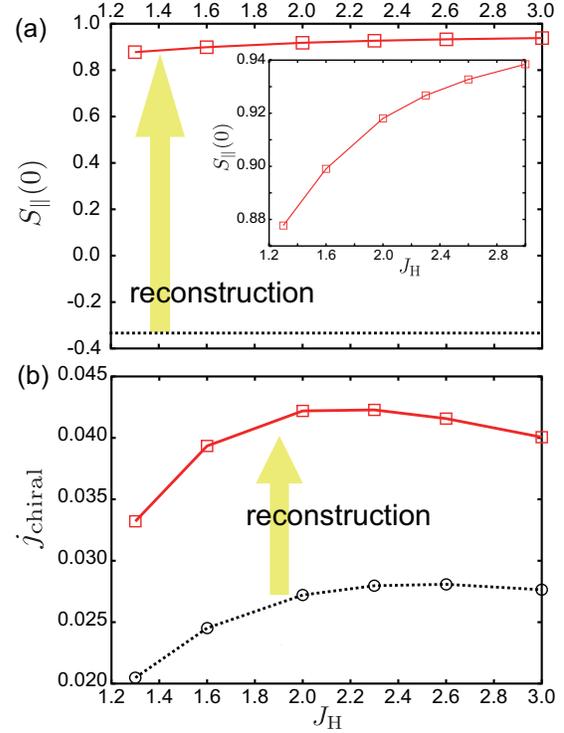}
\caption{
(Color online) 
The $J_{\rm H}$ dependence of 
(a) the spin correlation at the edge, $S_{\parallel}(0)$, and 
(b) the integrated edge current $j_{\mathrm{chiral}}$ for the optimized state. 
The dotted horizontal line in (a) represents the value of the spin correlation for the perfectly-ordered chiral state, -1/3.
The data connected by the dotted lines in (b) represent 
the integrated edge current for the perfectly-ordered chiral state.  
The inset in (a) shows the enlarged figure for the data for the optimized state.
}
\label{fig:J_dep}

\end{center}
\end{figure}

In Fig.~\ref{fig:J_dep}, we plot the $J_{\rm H}$ dependence of 
the spin correlation at the edge, $S_\parallel(0)$, and the integrated chiral edge current, $j_{\rm chiral}$, which 
is defined by the sum over a half of the system from one edge, 
\begin{equation}
j_{\rm chiral} = \sum_{i=0}^{L_x/2-1}j_\parallel(i).
\end{equation}
Figure~\ref{fig:J_dep}(a) shows that the ferromagnetic correlations are significantly developed at the edge.
As shown in the inset of Fig.~\ref{fig:J_dep}(a), $S_\parallel(0)$ is monotonically increased as $J_{\rm H}$ increases. 
On the other hand, as shown in Fig.~\ref{fig:J_dep}(b), 
although $j_{\rm chiral}$ shows a large enhancement up to almost twice by the reconstruction, 
it slightly decreases for larger $J_{\rm H}$. 
This is mainly comes from the suppression of edge current at the most outer layer at $i=0$ as discussed above. 

\begin{figure}[!h]
\begin{center}
\includegraphics[width = 8cm]{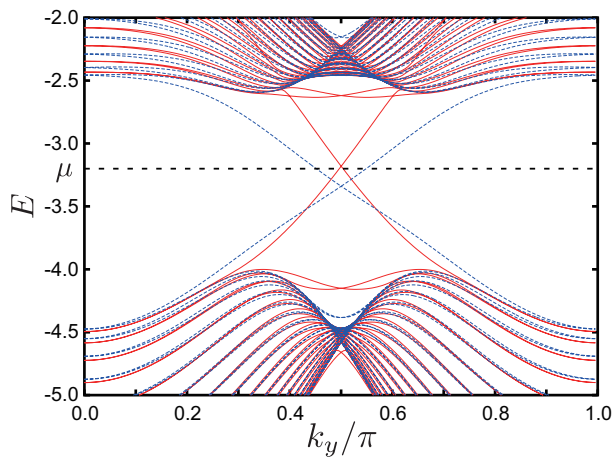}
\caption{
(Color online) 
Band structure for the optimized state (solid curves) and 
for the perfectly-ordered four-sublattice chiral state (dotted curves) near the chemical potential 
at 1/4 electron filling for $J_{\rm H}$ = 3.0.
The dashed horizontal line represents the chemical potential $\mu$ we set during the simulation. 
$k_y$ denotes the crystal momentum in the $y$ direction.
\label{fig:band}
}
\end{center}
\end{figure}

Let us discuss the enhancement of the chiral edge current from the electronic band structure. 
Figure~\ref{fig:band} shows the comparison of the band structures for the initial perfectly-ordered chiral state and for the optimized state. 
The bands are plotted as functions of the crystal momentum in the $y$ direction, $k_y$, in the energy region 
near the chemical potential at 1/4 filling for $J_{\rm H}=3.0$. 
There are chiral edge modes running across the bulk energy gap, for both magnetic structures,  
indicating that the topological nature characterized by the nonzero Chern number is intact by the reconstruction.
Remarkably, the edge modes are modified by the reconstruction to increase the Fermi velocity at the crossing point at $k_y=\pi/2$, 
which corresponds to the Fermi point at 1/4 filling. 
This is consistent with the increase of the total chiral edge current shown in Fig.~\ref{fig:sp_modu}(d). 

To summarize, we have numerically investigated the edge reconstruction of a magnetic Chern insulator 
with scalar chiral ordering realized in the Kondo lattice model on a triangular lattice. 
We have obtained the optimized spin configuration in the low-temperature limit in the system 
with open boundary condition in one direction by the Langevin-based simulation. 
As a result, we have found that ferromagnetic correlations are developed near the edge. 
Moreover, the persistent chiral edge current is enhanced; 
the integrated current for the optimized state becomes almost twice of that for the perfectly-ordered chiral order state. 
In the optimized state, the edge current induced by the magnetic reconstruction dominantly flows at inner layers, 
rather than the outmost edge of the system. 
The new current path is created at the boundary between the outmost ferromagnetic layer and the ordered bulk region. 
In other words, the system ``swallows" the edge state in the bulk by creating the ferromagnetic layer.

\begin{acknowledgment}
The authors thank K. Barros and Y. Kato for fruitful suggestions.
Y.A. is supported by Grant-in-Aid for JSPS Fellows.
This research was supported by Grants-in-Aid for Scientific Research (Grants No. 24340076 and 24740221),
the Strategic Programs for Innovative Research (SPIRE), MEXT, and the Computational Materials Science Initiative (CMSI), Japan.
\end{acknowledgment}

\bibliographystyle{jpsj}

\begin{thebibliography}{10}

\bibitem{PhysRevLett.49.405}
D.~J. Thouless, M.~Kohmoto, M.~P. Nightingale, and M.~den Nijs: Phys. Rev.
  Lett. {\bfseries 49} (1982) 405.

\bibitem{kohmoto1985topological}
M.~Kohmoto: Ann. Phys. (N.Y.) {\bfseries 160} (1985) 343.

\bibitem{PhysRevLett.61.2015}
F.~D.~M. Haldane: Phys. Rev. Lett. {\bfseries 61} (1988) 2015.

\bibitem{PhysRevB.62.R6065}
K.~Ohgushi, S.~Murakami, and N.~Nagaosa: Phys. Rev. B {\bfseries 62} (2000)
  R6065.

\bibitem{PhysRevLett.87.116801}
R.~Shindou and N.~Nagaosa: Phys. Rev. Lett. {\bfseries 87} (2001) 116801.

\bibitem{PhysRevLett.101.156402}
I.~Martin and C.~D. Batista: Phys. Rev. Lett. {\bfseries 101} (2008) 156402.

\bibitem{JPSJ.79.083711}
Y.~Akagi and Y.~Motome: J. Phys. Soc. Jpn. {\bfseries 79} (2010) 083711.

\bibitem{PhysRevLett.105.226403}
G.-W. Chern: Phys. Rev. Lett. {\bfseries 105} (2010) 226403.

\bibitem{PhysRevLett.109.166405}
J.~W.~F. Venderbos, M.~Daghofer, J.~van~den Brink, and S.~Kumar: Phys. Rev.
  Lett. {\bfseries 109} (2012) 166405.

\bibitem{PhysRev.100.675}
P.~W. Anderson and H.~Hasegawa: Phys. Rev. {\bfseries 100} (1955) 675.

\bibitem{PhysRevLett.81.1953}
J.-H. Park, E.~Vescovo, H.-J. Kim, C.~Kwon, R.~Ramesh, and T.~Venkatesan: Phys.
  Rev. Lett. {\bfseries 81} (1998) 1953.

\bibitem{PhysRevB.88.235101}
K.~Barros and Y.~Kato: Phys. Rev. B {\bfseries 88} (2013) 235101.

\bibitem{RevModPhys.78.275}
A.~Wei\ss{}e, G.~Wellein, A.~Alvermann, and H.~Fehske: Rev. Mod. Phys.
  {\bfseries 78} (2006) 275.

\bibitem{PhysRevLett.105.266405}
Y.~Kato, I.~Martin, and C.~D. Batista: Phys. Rev. Lett. {\bfseries 105} (2010)
  266405.

\end{thebibliography}

\end{document}